\documentclass[a4paper,11pt]{article}
\usepackage[utf8]{inputenc}
\usepackage[left=1.5cm,right=2cm,vmargin=2.5cm,footnotesep=0.5cm]{geometry}
\usepackage{amsmath}
\usepackage{amsfonts}
\usepackage{amssymb}
\usepackage{graphicx}
\usepackage{braket}
\usepackage{siunitx}
\usepackage{chemformula}
\usepackage{subcaption}
\usepackage{epstopdf}
\usepackage{multicol}
\usepackage{blindtext}
\usepackage{wrapfig}

\usepackage{setspace}
\usepackage{caption}
\makeatletter
\renewcommand{\maketitle}{
\bgroup\setlength{\parindent}{0pt}
\begin{flushleft}
  \textbf{\@title}

  \vspace{0.4cm}
  \@author
\end{flushleft}\egroup
}

\makeatother

\DeclareSIUnit{\rydberg}{Ry}
\DeclareSIUnit{\byte}{B}
\DeclareSIUnit\angstrom{\protect \text {Å}}

\newcommand{\bpm}{\begin{pmatrix}}
\newcommand{\epm}{\end{pmatrix}}
\newcommand{\eqname}{{Eq.}}

\newcommand{\Rcal}{\mathcal{R}}

\newcommand{\ft}{\text{f}}

\newcommand{\bR}{\boldsymbol{R}}

\newcommand{\bu}{\boldsymbol{u}}

\newcommand{\bPhi}{\boldsymbol{\Phi}}

\newcommand{\bUps}{\boldsymbol{\Upsilon}}

\newcommand{\bRcal}{\boldsymbol{\Rcal}}

\newcommand{\Avg}[2]{\left\langle #1\right\rangle_{#2}}

\renewcommand{\Im}{\mathrm{Im}}

\newcommand{\columnref}[1]{%
  \hyperref[#1]{%
    \pageref*{#1}-%
    \ifdim\zposx{#1}sp<.5\pdfpagewidth
      1% column 1
    \else
      2% column 2
    \fi
  }
}

\newcommand{\Vcal}{{\mathcal V}}

\begin{document}
\author{\textbf{\Large Lorenzo Monacelli$^1$$^\dagger$ and Nicola Marzari$^1$\\
\small$^1$Theory and Simulation of Materials (THEOS), and National Centre for Computational Design and Discovery of Novel Materials (MARVEL), École Polytechnique Fédérale de Lausanne, 1015 Lausanne, Switzerland\\
$^\dagger$lorenzo.monacelli@epfl.ch}}
\title{\Huge First-principles thermodynamics of \ch{CsSnI3}}
\maketitle
\section*{Abstract}
\begin{abstract}
\ch{CsSnI3} is a promising eco-friendly solution for energy harvesting technologies. It exists at room temperature in either a black perovskite polymorph or a yellow 1D double-chain, which irreversibly deteriorates in the air. In this work, we unveil the relative thermodynamic stability between the two structures with a first-principles sampling of the \ch{CsSnI3} finite-temperature phase diagram, discovering how it is driven by anomalously large quantum and anharmonic ionic fluctuations. Thanks to a comprehensive treatment of anharmonicity, the simulations deliver a remarkable agreement with known experimental data for the transition temperatures of the orthorhombic, rhombohedral, and cubic perovskite structures and the thermal expansion coefficient. We disclose how the perovskite polymorphs are the ground state above \SI{270}{\kelvin} and discover an abnormal decrease in heat capacity upon heating in the cubic black perovskite. Our results also significantly downplay the \ch{Cs^+} rattling modes' contribution to mechanical instability. The remarkable agreement with experiments validates our methodology, which can be systematically applied to all metal halides.
\end{abstract}
%\end{abstract}

\vspace{.5cm}
%\begin{multicols}{2}

\section*{Introduction}

Perovskites have an \ch{ABX3} chemical formula, where the B-site cation is octahedrally coordinated in a \ch{BX6} configuration and the \ch{A} cation sits within the cuboctahedral cavity formed by nearest-neighbor \ch{X} atoms in an \ch{AX12} polyhedron. Metal-halide perovskites (MHPs), in particular, are typically composed of a divalent B-site metal (e.g., \ch{Pb^2+}, \ch{Sn^2+}, \ch{Ge^2+}, \ch{Cu^2+}, \ch{Eu^2+}, and \ch{Ni^2+}) and monovalent A-site cation.
Inorganic MHPs usually employ \ch{Cs+} cations to improve stability.
Among all, the inorganic cesium lead halide \ch{CsPbI3} has been considered the best candidate for solar cells applications due to its suitable bandgap of \SI{1.73}{\electronvolt} and excellent optoelectronic properties\cite{Li2020}.
First reported in 2014, \ch{CsPbI3} perovskite solar cells (PSCs) have achieved remarkable progress in stability and power conversion efﬁciency through additive and composition engineering, interfacial modiﬁcations and optimization of the fabrication process\cite{Jiang2018}. Nevertheless, the presence of toxic lead hampers its deployment into general markets. \ch{CsSnI3} has established itself as the most promising eco-friendly alternative\cite{Wang2022}.

%Perovskite solar cells are one of the most promising technologies for energy harvesting. They are inexpensive and simple to manufacture but based, most commonly, on a hybrid organic-inorganic lead halide perovskite as the light-harvesting active layer\cite{Manser2016} (e.g., methylammonium lead halide and all-inorganic cesium lead halide).
%Despite their recent discovery for photovoltaic applications, they have already reached and surpassed the power conversion efficiency of standard silicon-based solar cells\cite{Min2021}, this latter achieved after more than sixty years of engineering and improvements.

%Whether PSCs can transition to large-scale industrial production depends on the possibility of overcoming some crucial drawbacks: their tendency to degrade under heat and humidity is detrimental for a product meant to spend hours in the sun, and the most efficient materials contain highly toxic lead. Research efforts are currently focused on boosting stability without sacrificing efficiency and delivering more environmentally benign solutions.

% Here the CsSnI3 introduction
\ch{CsSnI3} is polymorphic with two different phases coexisting at room temperature. The first black phase (B) is a standard perovskite crystal, which goes through three different phase transitions upon heating: it transforms from B-$\gamma$ (orthorhombic Pnma symmetry) to B-$\beta$ (tetragonal P4/mbm) at \SI{362}{\kelvin} and then to B-$\alpha$ (cubic Pm$\bar 3$m) at \SI{440}{\kelvin}\cite{Chung2012}. The second phase appears when \ch{CsSnI3} is exposed to air at room temperature; under these conditions, B-$\gamma$ transforms instead into a yellow phase (Y) with an orthorhombic Pnma space group, where the \ch{SnI6-} octahedra are connected into one-dimensional chains sharing one edge. In practice, \ch{CsSnI3} is synthesized at high temperature in the B-$\alpha$ phase\cite{Chung2012}. Then, when cooled to room temperature and exposed to air, it transforms into the yellow phase Y, suggesting that the perovskite structures (B-$\gamma$, B-$\beta$, and B-$\alpha$) are metastable under ambient conditions.

All the B phases display excellent optoelectronic properties and are considered the most promising eco-friendly candidates for high-performance PSCs. On the contrary, the Y phase is easily oxidized and irreversibly transforms to \ch{Cs2SnI6} whose absorption coefficient is ten times lower than the black perovskite polymorphs\cite{Lee2014,Karim2019}. Notably, also other isostructural tin metal-halides have been found to decompose into the Y phase\cite{Wang2022}. Therefore, understanding the mechanical stability between the perovskite polymorphs and the Y phase is a fundamental step to improving the overall stability of the \ch{CsSnI3} and its practical deployment.

Due to the experimental difficulties in the production of single crystals in the B-$\gamma$ phase, the structural characterization through X-ray spectroscopy has been achieved only recently\cite{Chung2012}, and our understanding of the B perovskite phase diagram is still in the early stages.
For example, the isomorph compound \ch{CsPbI3} was shown to form small domains of the orthorhombic black phase B-$\gamma$ within the cubic structure (B-$\alpha$) even at high temperatures\cite{Volonakis2016,Bertolotti2017}. So, it is not clear if the transition between ferroelectric B-$\gamma$, B-$\beta$ and paraelectric B-$\alpha$ in \ch{CsSnI3} is of the second-order displacive kind, where B-$\alpha$ is a high symmetry homogeneous crystal; or an order-disorder phase transition, where the crystal displays a local electric dipole even in the paraelectric phase\cite{Kotiuga2022}.
Moreover, theoretical calculations failed so far in reproducing even qualitatively the phase diagram of \ch{CsSnI3}, with the B-$\alpha$ phase predicted in many studies not to exist at any temperature\cite{Yu2013,daSilva2015}. This contrasts with experiments, which observe the B-$\alpha$ phase above \SI{440}{\kelvin}.
Further studies tried to include anharmonicity in the calculations, which is essential to describe the ferroelectric transition\cite{Patrick2015}, but the lack of algorithms to simulate lower symmetry phases, introduced only recently\cite{Monacelli2018}, prevented the simulation of the complete phase diagram. Moreover, the relative stability between Y and perovskite phases is extremely challenging for first-principles molecular dynamics, as the transition involves a macroscopic structural rearrangement of atoms. Thus, the phase diagram of \ch{CsSnI3} remains largely unknown: Are the transitions between B-$\gamma$, B-$\beta$, and B-$\alpha$ displacive or order-disorder? Is B-$\gamma$ or Y the lowest free energy phase at room temperature?

%The lead-halide inorganic perovskites forms small domains of the orthorhombic black phase B-$\gamma$ within the cubic structure (B-$\alpha$) even at high temperature, and these defects are argued to influence the photovoltaic performances\cite{Bertolotti2017}. The same possibility was also argued to occur in tin-halide structures\cite{Bechtel2018}. Evidence of a finite polarization in the high-temperature B-$\alpha$ phase was reported in \ch{CsSnBr3}\cite{Fabini2016}, further suggesting the presence of small distorted domains in the cubic phase.  

%This has important consequences on the performance of thermoelectric applications, as the proximity to a critical point of a displicative second-order phase transition strongly enhance the phonon-phonon scattering and provide a mechanism to increase the thermoelectric figure of merit without compromizing electronic properties, as it happens in tin-selenide, the highest thermoelectric figure of merit ever observed in a bulk material\cite{Aseginolaza2019}.

Here, we answer these questions by simulating the complete phase diagram of bulk \ch{CsSnI3} from first principles using state-of-the-art sampling techniques and disclosing the origins of the formation of the Y phase. %Furthermore, we employ the stochastic self-consistent harmonic approximation (SSCHA)\cite{SSCHA} to account for the much-needed anharmonic quantum and thermal fluctuations. 
In the process, we both elucidate the displacive character of the ferroelectric phase transitions in the black perovskite, highlight its anomalous heat capacity, and show the impact of the tin and cesium rattling motions on mechanical stability.

\section*{Results}

The importance of anharmonicity has been recently discovered in the isostructural compounds \ch{CsPbI3}\cite{Marronnier2018,Kaiser2021} and \ch{CsPbBr3}\cite{Tadano2022}. As will be seen below, anharmonicity in thermal and quantum ionic fluctuations plays a crucial role in the thermodynamic properties and phase diagram of \ch{CsSnI3}. We account for this by employing the stochastic self-consistent harmonic approximation (SSCHA)\cite{Monacelli2021}, combined with density-functional theory (DFT) at the PBEsol level\cite{PBEsol}. The SSCHA, essential to this work, captures the strongly anharmonic fluctuations of the ions by optimizing a nuclear quantum distribution that minimizes the free energy\cite{Errea2014}. Within the SSCHA, one can optimize the average ionic positions (the centroids of the nuclear quantum distribution), the lattice vectors and cell volume as a function of temperature. The method is stochastic and samples the energy landscape extracting configurations with randomly displaced ions and evaluating the respective energies and forces within DFT. The advantages of the SSCHA compared to other state-of-art approaches, like \emph{ab initio} molecular dynamics, are the direct access to free energies, also exploiting symmetry constraints, and the inclusion of quantum nuclear zero-point motion. Moreover, unlike other approximate methods like the time-dependent energy landscape (TDEP\cite{TDEP}), it is nonempirical and has no internal free parameters that could affect the results (like the choice of the diagrams to include in the phonon self-energy or the order of the fit for the energy landscape).

\begin{figure*}
    \centering
    \includegraphics[width=\textwidth]{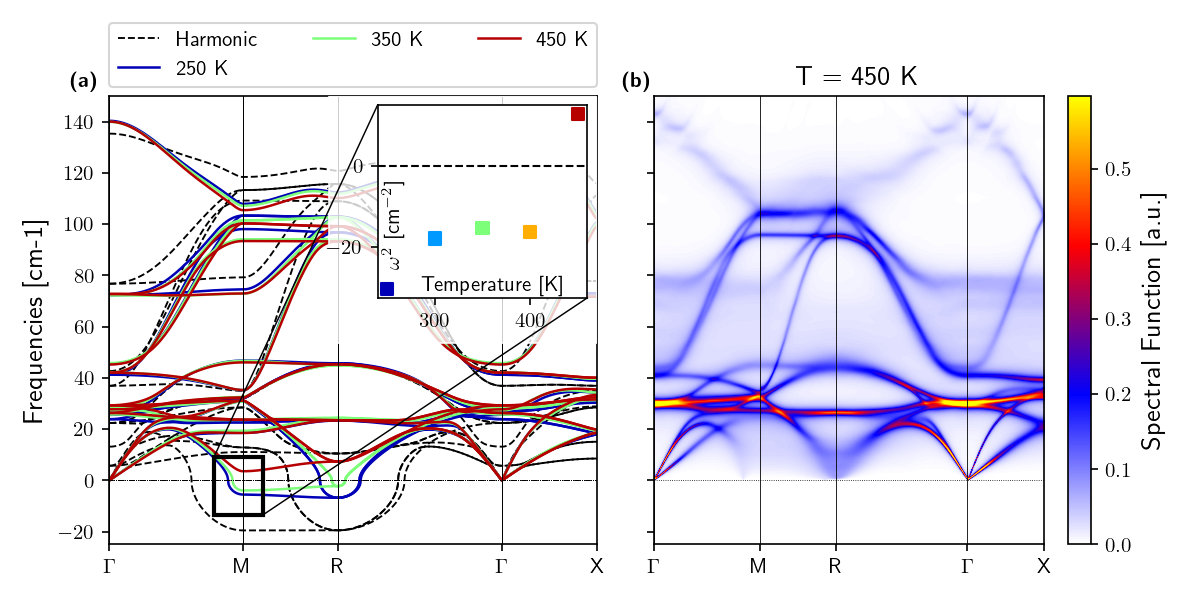}
    \caption{\textbf{(a)}: Phonon dispersions of the B-$\alpha$ phase computed within the harmonic approximation (black dashed lines) contrasted with the full inclusion of anharmonicity within the SSCHA at different temperatures; negative values indicate unstable (imaginary) vibrations. In the inset, we report the second derivative of the vibrational free energy for the ionic displacement at M that transforms B-$\alpha$ into B-$\beta$. \textbf{(b)} Vibrational spectrum of B-$\alpha$ at \SI{450}{\kelvin}. The finite width of the bands is given by physical lifetimes due to phonon-phonon scattering. 
    Most phonon bands have extraordinary linewidths leading to coherent thermal transport across different bands\cite{Simoncelli2019}.}
    \label{fig:hessian:spec}
\end{figure*}

The most simple structure for \ch{CsSnI3} is the standard cubic perovskite B-$\alpha$ (space group Pm$\bar 3$m), with five atoms in the primitive cell. Despite its geometrical simplicity, it is a saddle-point of the Born-Oppenheimer energy landscape, and the harmonic phonon dispersion presents imaginary frequencies (\figurename~\ref{fig:hessian:spec}\textbf{a}). Moreover, extrinsic thermal effects (e.g., volume expansion) further destabilize the structure, introducing an additional imaginary mode at $\Gamma$\cite{daSilva2015}.
To assess the stability of the B-$\alpha$ phase, we computed the Hessian of the free energy with respect to the centroids\cite{Bianco2017_structural}, which defines an effective anharmonic temperature-dependent static phonon dispersion.
The critical temperature at which B-$\alpha$ becomes mechanically stable occurs when the phonon dispersion becomes positive in the whole Brillouin zone (\figurename~\ref{fig:hessian:spec}\textbf{a}); we report more details on the calculations in the Methods section. 
An ionic displacement with imaginary frequency in the M-R region of the Brillouin zone identifies the low-temperature B-$\alpha$ instability. The instability disappears at R between \SI{350}{\kelvin} and \SI{400}{\kelvin}, then at M
at \SI{450}{\kelvin} as B-$\alpha$ becomes stable.
This is in very good agreement with experiments that show how the B-$\beta$ phase transforms into the B-$\alpha$ between \SI{430}{\kelvin} and \SI{440}{\kelvin}\cite{Chung2012}.
At variance with the unstable M-R phonon modes, other frequencies display only a slight variation with temperature, despite their substantial difference with respect to the harmonic spectrum. Such a strongly anharmonic character becomes even more apparent when examining the vibrational spectrum of the B-$\alpha$ phase. This is reported in \figurename~\ref{fig:hessian:spec}(\textbf{b}) and is evaluated as the trace of positions auto-correlation functions within the time-dependent self-consistent harmonic approximation (TD-SCHA)\cite{Monacelli2021} (more details in Methods); the very broad linewidth of the phonon bands points to their short lifetimes. The phonon-phonon scattering is so strong that the character of the dispersion disappears, and almost all phonons merge. 
This justifies a posteriori the necessity of dealing carefully with such strongly anharmonic crystals, and the overlap between different phonon bands points toward the importance of coherences and Wigner transport for thermal conductivity\cite{Simoncelli2019}, neglected in the standard Boltzmann theory, and the necessity to account for the overdamped regime of the low-frequency modes.%, which cannot be employed in metal halide perovskite.

\begin{figure*}[bt]
    \centering
    \includegraphics[width=0.7\textwidth]{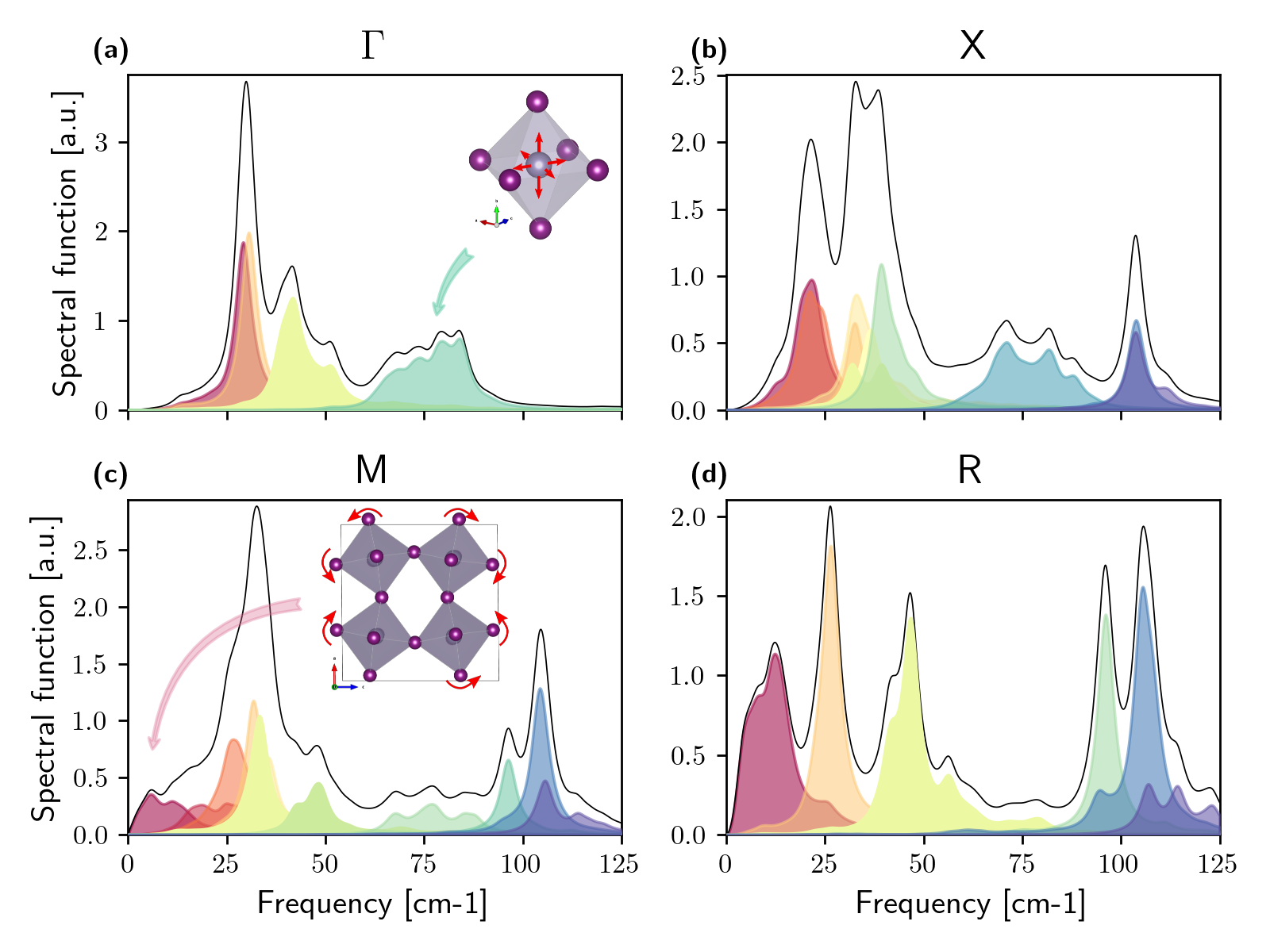}
    \caption{Spectral function of the B-$\alpha$ phase at \SI{450}{\kelvin} and different high-symmetry points. The simulations are performed with a smearing of \SI{5}{\per\centi\meter}; thus, the overall shapes of the peaks represent the intrinsic finite lifetimes due to phonon scattering. We report the single contribution of each phonon mode, highlighted with different colors. In panel \textbf{(a)} we neglected the LO-TO splitting at $\Gamma$. % to simulate the self-doped metallic behavior induced by tin vacancies. 
    Atomic vibrations for tin rattling are reported in the inset in \textbf{(a)}. The \ch{SnI6} octahedra tilting driving the phase transition from B-$\alpha$ to B-$\beta$ is reported in \textbf{(c)}; a similar tilting is also present in R \textbf{(d)} at very low frequencies.}
    \label{fig:spectral}
\end{figure*}

To further investigate the vibrational properties of the B-$\alpha$ phase, we dissected the spectral function separating the contribution of each mode in the $\Gamma$, M, R, and X high-symmetry points in \figurename~\ref{fig:spectral}, including also the effects of four-phonon scattering self-consistently\cite{Monacelli2021} (see Methods).

%This, joined with the dominant role of high-order anharmonicity and the non-Lorentzian behavior of spectral functions of phonons, makes the calculation of thermal transport in metal halide perovskites an open challenge that requires the development of new methodologies.

%In this calculation, we include anharmonic phonon-phonon scattering up to any order in the self-energy and explicitly consider the 4-phonon scattering processes, thanks to the revolutionary Lanczos algorithm devised for the TD-SCHA\cite{Monacelli2021}.

Almost all phonon modes in the Brillouin zone display a peak shape departing from the standard Lorentzian.
The mode which drives the phase transition between B-$\alpha$ and B-$\beta$ is shown as a dark-red broad band at $M$ and $R$ (\figurename~\ref{fig:spectral}\textbf{c},\textbf{d}). This band represents the tilting of the \ch{SnI6-} octahedra along the different directions; it has a broad featureless spectrum spanning low frequencies up to \SI{20}{\per\centi\meter} and is strongly over-damped with a lifetime shorter than the oscillation period. A similarly broad spectrum has already been measured in the isostructural \ch{CsPbBr3}\cite{LaniganAtkins2021}. However, even higher energy modes show nontrivial peak shapes; e.g., the phonon branch around \SI{75}{\per\centi\meter} at $\Gamma$, $X$, and $M$ has a large broadening of \SI{30}{\per\centi\meter}. This is the mode for \ch{Sn} rattling inside the \ch{SnI6} cages, underlining how temperature delocalizes the bonding of tin (\figurename~\ref{fig:spectral}\textbf{a},\textbf{b}\textbf{c}). 
%Moreover, the highest frequency purple peak at R shows broad satellite features (\figurename~\ref{fig:spectral}\textbf{d}). It represents the breathing of the $\ch{SnI6}$ cages, where the \ch{SnI6} octahedra in near neighbor cells grow and contract.
In contrast, the modes involving the motion of \ch{Cs+} ions have a Lorentzian shape (e.g., the one marked in red around \SI{31}{\per\centi\meter} at $\Gamma$, or the one in orange around \SI{26}{\per\centi\meter} at R; \figurename~\ref{fig:spectral}\textbf{a,d}); this is a signature that atoms vibrate similarly to an \emph{effective} harmonic oscillation. 

These observations challenge the common assumption that \ch{Cs+} ions rattle inside the over-sized octahedra of the perovskite structure\cite{Chung2012,Xie2020} and that this motion plays a crucial role in the stability of the perovskite structure, an assumption incorrectly corroborated by quasi-harmonic simulations, which show that \ch{Cs^+} vibrations become imaginary at the $\Gamma$ point upon volume dilation\cite{daSilva2015,Xie2020}. Indeed, \figurename~\ref{fig:hessian:spec}\textbf{(a)} shows a relevant frequency shift with respect to the Harmonic value (from \SI{5}{\per\centi\meter} to \SI{28}{\per\centi\meter} at \SI{450}{\kelvin}); however, 
\ch{Cs+} motion is stable already at \SI{250}{\kelvin} when intrinsic anharmonicity is accounted for and barely depends on temperature.
Despite the agreement between our simulations showing the B-$\alpha$ phase becoming stable between \SI{400}{\kelvin} and \SI{450}{\kelvin} and the experimental transition temperature, questions remain regarding the order of the phase transition (first or second) and its character (order-disorder or displacive).
To answer these questions, we further investigated the ferroelectric transition in the B-$\gamma$ and B-$\beta$ phases: we prepared the starting trial nuclear density matrix in the orthorhombic B-$\gamma$ phase and optimized the SSCHA distribution, including the cell shape and centroids within the symmetry constraints of the Pnma orthorhombic group from \SI{250}{\kelvin} to \SI{450}{\kelvin} (each \SI{50}{\kelvin}). The procedure is repeated with the symmetry constraints of B-$\beta$ (P4/mbm).% and reported in \figurename~\ref{fig:phase:diagram}.
\begin{figure*}[tb]
  \centering
  \includegraphics[width=0.7\textwidth]{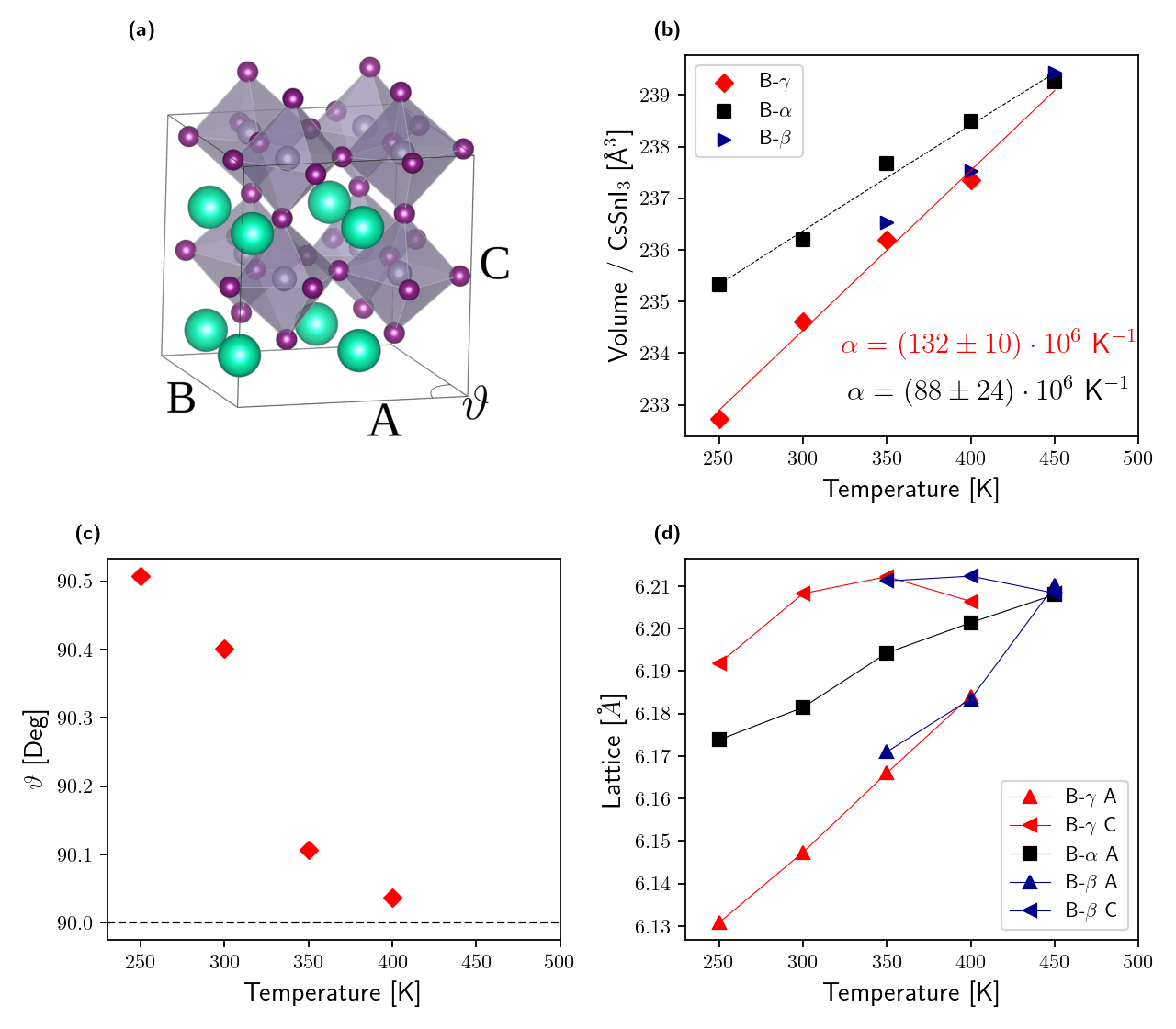}
  \caption{Structure of the black perovskite \ch{CsSnI3}.
  \textbf{(a)} The conventional cell of 48 atoms commensurate with both B-$\alpha$, B-$\beta$ and B-$\gamma$ (rendered with VESTA\cite{VESTA}). For cubic B-$\alpha$, it is a 2x2x2 supercell. The primitive unit cell of B-$\beta$ and B-$\gamma$ is shown in red, identified by $a$ and $b$ and $C$ segments. The $\vartheta$ angle is $90^\circ$ when $a=b$: in the tetragonal B-$\beta$. When also $A=C$, the phase is cubic, as in B-$\alpha$. \textbf{(b)} Volume expansion as a function of temperature; the thermal expansion coefficient $\alpha_v$ computed at \SI{300}{\kelvin} is reported on the plot. \textbf{(c)} $\vartheta$ angle between $A$ and $B$;
  \textbf{(d)} Size of the lattice parameters. 
  Analyzing the lattice parameters $\vartheta$, $A$ and $C$, we conclude that the structure transit into the rhombohedral B-$\beta$ between \SI{350}{\kelvin} and \SI{400}{\kelvin} and into the cubic one B-$\alpha$ at \SI{450}{\kelvin}. These transitions are confirmed by further symmetry analysis of the centroids.} 
  \label{fig:phase:diagram}
\end{figure*}
To detect the transition, we measure the distortion of the conventional cell with 48 atoms (commensurate with all the three phases,  \figurename~\ref{fig:phase:diagram}\textbf{a}) in analogy with X-ray diffraction experiments. Namely, the orthorhombic to rhombohedral transition is identified by the $\vartheta$ angle between the lattice vectors of the almost cubic 48-atom supercell  (\figurename~\ref{fig:phase:diagram}\textbf{c}) and the rhombohedral to cubic transition by the relative size of the lattice parameters $A$ and $C$ (\figurename~\ref{fig:phase:diagram}\textbf{d}). The deviation of $\vartheta$ from $90^\circ$ (\figurename~\ref{fig:phase:diagram}\textbf{c}) shows how the B-$\gamma$ phase transforms continuously into B-$\beta$ between \SI{350}{\kelvin} and \SI{400}{\kelvin}. The transition to the cubic B-$\alpha$ phase occurs at around \SI{450}{\kelvin}, as shown by the value at which the cell becomes cubic (\figurename~\ref{fig:phase:diagram}\textbf{d}) and the temperature at which the volumes of the simulations constrained along the B-$\gamma$ and B-$\beta$ phases intersect the one of the B-$\alpha$ phase (\figurename~\ref{fig:phase:diagram}\textbf{b}). A symmetry analysis of the centroids confirms both transitions. The resulting phase diagram is in excellent agreement with experimental data (B-$\beta$ at \SI{362}{\kelvin} and B-$\alpha$ at \SI{440}{\kelvin}\cite{Chung2012}): the match between this simulation and the stability analysis of the phase B-$\alpha$ indicates that there is no metastability region for the B-$\alpha$ phase (no hysteresis between B-$\beta$ and B-$\alpha$), and it confirms a second-order phase transition supporting the displacive scenario, as the crystal has no local ferroelectricity above the critical temperature.

\ch{CsSnI3} displays a remarkably large thermal expansion coefficient $\alpha_v = \frac{1}{V}\left( \frac{\partial V}{\partial T}\right)_P$ at room temperature of \SI{94(28)e6}{\per\kelvin} in the cubic B-$\alpha$ phase and \SI{132(10)e6}{\per\kelvin} in the orthorhombic B-$\gamma$ phase. This latter is in excellent agreement with \emph{in situ} diffraction experiments (\SI{126e6}{\per\kelvin} in the B-$\gamma$ phase\cite{Chung2012}), validating the accuracy of the PBEsol functional and our treatment of anharmonicity.
These values are uncommon when compared with other materials: for example, the isostructural compound \ch{MgSiO3} has a $\alpha_v$ equal to \SI{15e6}{\per\kelvin}\cite{Wang1991}, while \ch{SrZrO3} and \ch{BaZrO3} \SI{29.8e6}{\per\kelvin} and \SI{10.6e6}{\per\kelvin} respectively\cite{Zhao1991}. 
The volumetric thermal expansion coefficient $\alpha_v$ of B-$\gamma$ \ch{CsSnI3} is the largest known for a crystal, approaching those of amorphous systems and liquids\cite{Chung2012}.

%To point out the critical temperature for the transition into the cubic phase, we simulated the free energy difference of the three black polymorphs B-$\alpha$, B-$\beta$ and B-$\gamma$ phases (\figurename~\ref{fig:yellow:pd}\textbf{a}).
%In a displicative second-order phase transition, the transition matches the critical temperature below which the high-symmetry phase becomes unstable. Indeed, above \SI{300}{\kelvin}, the free energy of all the phases of perovskites are almost degenerate (below \SI{2}{\milli\electronvolt} per formula unit) and fall within the stochastic error. To dissect the critical temperature of B-$\alpha$ precisely, we computed the curvature of the free energy landscape with respect to the order parameter of the transformation. At the critical point, the curvature transit from negative (unstable) to positive (stable) (\figurename~\ref{fig:yellow:pd}\textbf{b}). The cubic B-$\alpha$ is found stable already at \SI{400}{\kelvin}; therefore, the cubic B-$\alpha$ and the rhombohedral B-$\beta$ phase can coexist in a narrow range of temperatures (up to \SI{450}{\kelvin}).
%This behavior identifies a week first-order transition.

%The coexistence between B-$\beta$ and B-$\alpha$ between \SI{400}{\kelvin} and \SI{450}{\kelvin} supports the recent observation of ferroelectric activity even at high temperature in the B-$\alpha$ of \ch{CsPbX3}\cite{Bertolotti2017}, indicating that \ch{CsSnI3} behaves similarly.

%\subsection*{Spectral properties}

\begin{figure*}[tb]
  \centering
  \includegraphics[width=0.8\textwidth]{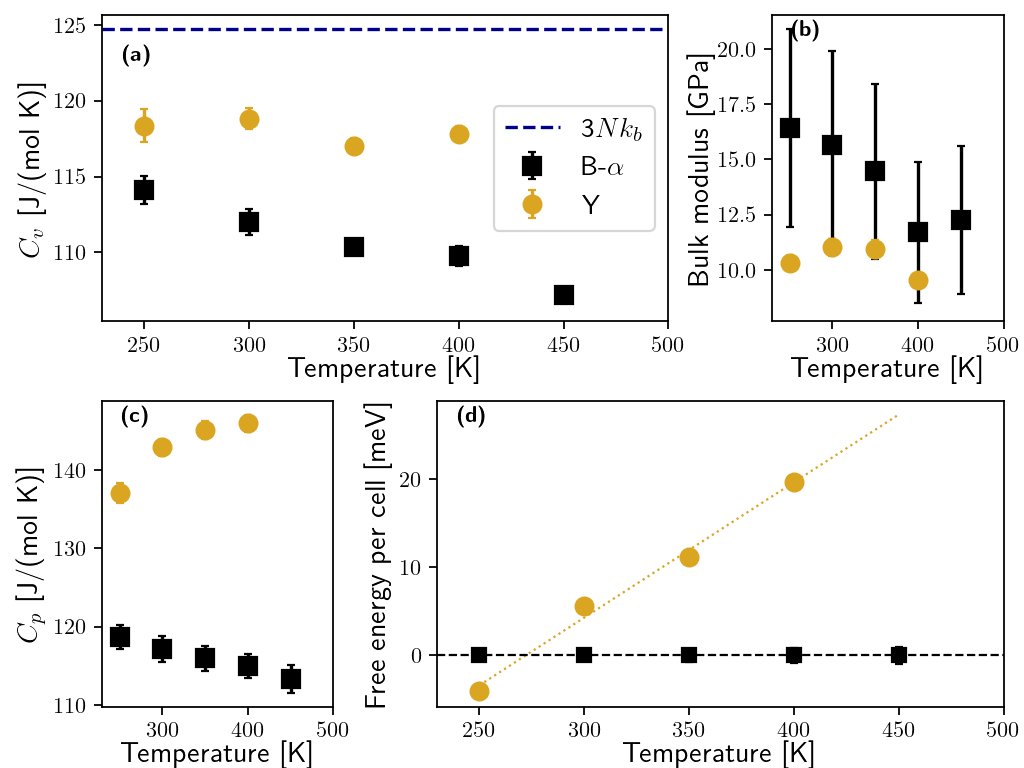}
  \caption{Thermodynamic properties of the B-$\alpha$ and Y phases of \ch{CsSnI3}. \textbf{(a)} Constant-volume heat capacity. In a harmonic system, $C_v$ should approach at high temperatures the classical value $3N k_b$ (\SI{124.7}{\joule\per\mole\per\kelvin}), reported here as a blue dashed line. However, B-$\alpha$ shows an anomalous heat capacity decrease upon heating, and also Y does not converge to the expected classical value. \textbf{(b)} Bulk modulus.  \textbf{(c)} Constant-pressure heat capacity.
  \textbf{(d)}
Free energy difference between the B-$\alpha$ and Y phases. Above about \SI{270}{\kelvin}, the B-$\alpha$ phase becomes favored. Here, B-$\alpha$ is used as a prototype also for B-$\beta$ and B-$\gamma$, as their free energy differences are below \SI{2}{\milli\electronvolt} per formula unit in the whole temperature range studied.}
  \label{fig:yellow:pd}
\end{figure*}

Using the SSCHA we can compute the free energy at finite temperatures even in materials with strong intrinsic anharmonicity, as is the case for the B-$\alpha$ phase of \ch{CsSnI3}. To shed light on the decomposition of this black perovskite into the orthorhombic yellow phase (Y) when the sample is exposed to air, we run a new SSCHA calculation on the Y phase and compared its free energy and thermodynamic properties as a function of temperature with that of the B-$\alpha$ phase (\figurename~\ref{fig:yellow:pd}). Since the B-$\alpha$, B-$\beta$ and B-$\gamma$ phases transform through second-order phase transitions, their free energy differences are below \SI{2}{\milli\electronvolt} per formula unit in the temperature range studied (\SI{250}{\kelvin}-\SI{450}{\kelvin}). Therefore, we employed the B-$\alpha$ as a prototype for the all three black perovskites when assessing its relative stability with respect to the Y phase, as the SSCHA can reach a lower stochastic error and a better thermolimit extrapolation exploiting the higher number of symmetries of the cubic phase.

%\subsection*{Thermodynamics}

%The SSCHA simulations give access to thermodynamic potentials and their derivatives, from which we evaluate the thermodynamic properties of the B-$\alpha$ and $Y$ phases.
We compare in \figurename~\ref{fig:yellow:pd}\textbf{(a,b,c)} the constant-volume heat capacity $C_v$, the bulk modulus, and constant-pressure heat capacity $C_p$ for the B-$\alpha$ and Y phases. Notably, as a function of temperature, the constant volume heat capacity (\figurename~\ref{fig:yellow:pd}\textbf{a}) of the B-$\alpha$ decreases anomalously. %The temperature range of the study lies above the Debye temperature of the highest phonon mode (\SI{230}{\kelvin}) and, 
According to harmonic theory, above the Debye temperature $T_D \approx \SI{230}{\kelvin}$ $C_v$ should reach $3Nk_b=\SI{124.7}{\joule\per\mole\per\kelvin}$ (blue dashed line in \figurename~\ref{fig:yellow:pd}\textbf{a}). 
The anomalous thermal dependence of the heat capacity in the B-$\alpha$ phase instead further underlines the anharmonicity of the crystal. In fact, according to the SSCHA (see SI, sec.~\ref{sec:specific:heat}), the contribution of each phonon to the heat capacity is:
\begin{equation}
    C_v(T) \stackrel{T\gg T_D}{=}k_b\sum_{\mu=1}^{3N} \left[ 1 + \frac{T}{\omega_\mu} \left(\frac{\partial \omega_\mu}{\partial T}\right)_V\right],
    \label{eq:heat:sscha}
\end{equation}
where the first term is the standard Dulong-Petit model for solids ($3Nk_b$), while the second one accounts for the intrinsic anharmonic shift of frequencies with temperature at constant volume; this is not captured by the quasi-harmonic approximation. The B-$\alpha$ phase instability close to \SI{450}{\kelvin} (\figurename~\ref{fig:hessian:spec}\textbf{a}) generates a softening of the full phonon branch between M and R, which, thanks to the $1/\omega$ factor, enhances the effect of anharmonicity and explains the negative slope of the heat capacity before the phase transition.

The constant-volume heat capacity of the Y phase does not show the same anomaly: it is almost independent of temperature, and it deviates from the value predicted by the harmonic theory by 6\%. 
The difference between the Y and B-$\alpha$ phases is further enhanced at constant pressure (\figurename~\ref{fig:yellow:pd}\textbf{c}), due to the higher thermal expansion coefficient of the Y phase ($\alpha_v = \SI{117(4)e6}{\per\kelvin}$).
The bulk moduli of both phases are very similar, with a softer value for the Y phase at low temperature. The bulk modulus of the cubic perovskite structure shows a slight decrease with temperature and its much smaller value when compared with other perovskite structures (e.g., $B = \SI{170}{\giga\pascal}$ in \ch{SrTiO3}, about than 15 times larger) is at the root of the remarkable softness of \ch{CsSnI3} and its sizable thermal expansion coefficient.

%Even if the B-$\alpha$ phase has a critical phase transition between \SI{400}{\kelvin} and \SI{450}{\kelvin}, the logarithmic divergence in the specific heats is invisible in the scale of our simulations, sampled every \SI{50}{\kelvin}.

The free energy calculations unveil how the B-$\alpha$ phase is more stable than the Y above \SI{270}{\kelvin} (\figurename~\ref{fig:yellow:pd}\textbf{(d)}); also note that even if the B-$\gamma$ phase is more stable at that temperature, its free energy difference with the B-$\alpha$ is negligible  (\SI{2}{\milli\electronvolt} per formula unit). This result apparently challenges experiments showing a spontaneous transformation of the B-$\gamma$ into the Y phase at room temperature.
However, such deterioration of the black perovskite has been observed only in samples exposed to air\cite{Wang2022}. Since \ch{CsSnI3} is synthesized as a powder, surface effects can be very sizable, and contamination of the sample with water and oxygen could alter the relative stability between the two phases. Moreover, it is known that the Y phase is easily oxidized and irreversibly transforms into the \ch{Cs2SnI6}\cite{Lee2014,Karim2019}. Therefore, increasing the volume/surface ratio of the material (i.e., by growing larger crystals) would be a promising route to prevent the formation of the Y phase in the first place. The steep increase in the free energy difference between the two phases also shows how heating could efficiently remove contamination of the Y phase inside the solar cell. 
In conclusion, we simulated the complete ambient pressure phase-diagram of \ch{CsSnI3}, showing an excellent agreement within 15 \% with the experimental transition temperatures for the black perovskite structures; this accuracy is comparable to the one of SSCHA+DFT (at the PBEsol level) in other materials where ionic fluctuations drive the phase transition rather than electronic processes\cite{Bianco2019NbS2,Diego2021,Monacelli2022quantum_arxiv}, and achieved only through a complete treatment of anharmonicity. The lower free energy of the black perovskite structure compared to the yellow phase at room temperature is beneficial for the stability of the system, as preventing the formation of the yellow phase is an important step toward the stabilization of the perovskite structure in the air.
Our approach is general and can be employed in any other metal halides, paving the way to a reliable high-throughput study of these materials with first-principles simulations.

\section*{Acknowledgements}
LM acknowledges PRACE for awarding access to Joliot-Curie Rome at TGCC, France; CINECA under the ISCRA initiative for access to MARCONI100, Italy; and CSCS, for sharing the resources of the hybrid partition of Piz Daint, Switzerland (project IDs c29 and s1139). This project was founded by European Union under the Marie Curie Fellowship (project codename THERMOH).

\newpage

\section*{Methods}

We studied the structure and the electronic properties of \ch{CsSnI3} within density-functional theory (DFT) in the PBEsol approximation\cite{PBEsol}, using the Quantum ESPRESSO distribution \cite{Espresso} employing a plane-wave basis set, norm-conserving pseudopotentials from the Pseudo-Dojo library\cite{Dojo}, and a cutoff of 70 Ry. The Brillouin zone for electrons is sampled with an 8x8x8 uniform mesh with respect to the primitive-reciprocal cell of the B-$\alpha$ structure; this sampling is appropriately re-scaled in the other phases.
%We employed the exchange-correlation functional PBEsol to compute stress, forces, and energy.

Anharmonicity and phonons are studied with the Stochastic Self-Consistent Harmonic Approximation (SSCHA)\cite{Monacelli2018,Monacelli2021}. SSCHA calculations are performed on a 2x2x2 supercell of the B-$\alpha$ phase containing 40 atoms; their convergence has been verified by comparing the results obtained in a 3x3x3 supercell with 135 atoms at one temperature (\SI{300}{\kelvin}), showing no significant differences. To converge the free energy and the thermodynamic properties (heat capacity and bulk modulus) with the supercell, we exploited the natural division of the SSCHA free energy into a long-range harmonic-like term and the short-range anharmonic correction\cite{Monacelli2021}. The harmonic-like free energy has been interpolated into an 8x8x8 supercell containing 2560 atoms, accounting also for long-range electrostatic interactions (LO-TO splitting). The same conditions were also applied to the Y phases.

To evaluate the thermodynamic properties and second derivatives of the free energy within the SSCHA, we introduce a new algorithm. The properties of interest for this work are related to entropy and pressure as:
\begin{equation}
    C_v = T\left(\frac{\partial S}{\partial T}\right)_V 
    \label{eq:cv}
\end{equation}
\begin{equation}
    \frac{\alpha_v}{\beta_T} = \left(\frac{\partial P}{\partial T}\right)_V 
    \label{eq:bt}
\end{equation}
\begin{equation}
    C_p = C_v + VT \alpha_v \left(\frac{\partial P}{\partial T}\right)_V
    \label{eq:cp}
\end{equation}
where $\beta_T$ is the isothermal compressibility (the inverse of the bulk modulus) and $\alpha_v$ is the volumetric expansion coefficient, shown in \figurename~\ref{fig:phase:diagram}.
Thanks to correlated sampling\cite{Errea2014,SSCHA,Miotto2018,TOLOMEO}, we can slightly vary the temperature at a fixed volume without the need for any new DFT calculation, obtaining the free energy and its derivatives (the entropy $S$ and the pressure $P$) at temperatures surrounding the simulated one. We estimate the thermodynamic relations in Eqs.~\eqref{eq:cv},\eqref{eq:bt},\eqref{eq:cp} by employing a finite-difference approach on the correlated sampling simulations (see SI for more details).

The free energy Hessian needed to study the stability of the B-$\alpha$ phase reported in \figurename~\ref{fig:hessian:spec} is computed as
\begin{equation}
    \frac{d^2F}{d\Rcal d\Rcal} = \stackrel{(2)}{\bPhi} - \frac 12 \stackrel{(3)}{\bPhi} \left(1 + \frac 12{\boldsymbol{\chi}}\stackrel{(4)}{\bPhi}\right)^{-1}{\boldsymbol{\chi}}\stackrel{(3)}{\bPhi},
    \label{eq:hessian}
\end{equation}
where 
\begin{equation}
    \stackrel{(2)}{\bPhi}_{ab} = \left<\frac{d^2V}{dR_adR_b}\right> \qquad
    \stackrel{(3)}{\bPhi}_{ab} = \left<\frac{d^3V}{dR_adR_bdR_c}\right>\qquad 
    \stackrel{(4)}{\bPhi}_{abcd} = \left<\frac{d^4V}{dR_adR_bdR_cdR_d}\right>,
\end{equation}
and $\boldsymbol{\chi}$ is the two-phonon free propagator as described in Refs.\cite{Bianco2017_structural,Monacelli2018,SSCHA}. Usually, four-phonon scattering processes can be neglected in the inversion of \eqname~\eqref{eq:hessian}\cite{Bianco2017_structural,Diego2021,Bianco2018_H3SPRB,Bianco2019NbS2,Aseginolaza2019,Aseginolaza2019_SnS,Errea2020_Nature} as it plays a negligible effect (bubble approximation). However, we found that its inclusion is necessary here to describe the subtle effects involved in the phase transition; otherwise, the B-$\alpha$ becomes unstable at all temperatures. 

%\subsection*{Calculation of the spectral properties}
The spectral function reported in \figurename~\ref{fig:hessian:spec} and \ref{fig:spectral} is defined as:
\begin{equation}
    \sigma(q, \omega) = -\frac{\omega}{\pi} \sum_\mu\Im G_\mu(q, \omega),
\end{equation}
where $\Im G_\mu(q, \omega)$ is the imaginary part of the $\mu$-th phonon dynamical Green function at $q$, evaluated with the time-dependent self-consistent harmonic approximation (TD-SCHA) nonperturbatively\cite{Monacelli2021}. The $\omega/\pi$ factor makes the integral proportional to the total number of modes.

In \figurename~\ref{fig:spectral} we report the full spectral function including the $\stackrel{(4)}{\bPhi}$ term in the self-energy evaluated on the 40-atom supercell; this is made possible by the use of the Lanczos algorithm recently introduced\cite{Monacelli2021}. This cell is sufficient to get converged spectral function due to the short lifetime of phonons in this material. We choose the value of the smearing by checking the convergence with the supercell, possible only when neglecting the four-phonon scattering $\stackrel{(4)}{\bPhi}$ in the self-energy with the interpolation introduced in Ref.\cite{Bianco2018_H3SPRB}.

	\bibliographystyle{ieeetr}
	\bibliography{biblio}
%\end{multicols}

	\newpage
\appendix
    \section{Heat capacity within the SSCHA theory}
\label{sec:specific:heat}
The heat capacity is the derivative of entropy with respect to temperature:
\begin{equation}
    C_v = T \left(\frac{dS}{dT}\right)_v;
\end{equation}
since the entropy is the derivative of the free energy
\begin{equation}
    S = -\left(\frac{dF}{dT}\right)_v,
\end{equation}
the heat capacity is proportional to the second derivative of the free energy with respect to temperature. 

The entropy within the SSCHA is the entropy of the auxiliary harmonic Hamiltonian, i.e.
\begin{equation}
    S(T) = k_b\sum_{\mu = 1}^{3N}\left[ -\ln(1 - e^{-\beta\omega(T)}) + \beta\omega(T)\frac{1}{e^{\beta\omega(T)} - 1}\right],
\end{equation}
where $\omega_\mu(T)$ are the frequencies of the auxiliary harmonic Hamiltonian that solves the SSCHA at each temperature.
The heat capacity is therefore
\begin{equation}
    C_v = T\frac{dS}{dT} = T\frac{\partial S}{\partial T} + T\sum_{\mu = 1}^{3N} \frac{\partial S}{\partial\omega_\mu}\frac{\partial \omega_\mu}{\partial T} =  C_v^{\text{(harm)}} +  C_v^{\text{(anharm)}}.
\end{equation}
The first term is the standard specific heat of the quantum harmonic oscillator:
\begin{equation}
    C_v^{\text{(harm)}} = T \frac{\partial S}{\partial T} = k_b\sum_{\mu=1}^{3N}\frac{\beta^2\omega_\mu^2e^{\beta\omega_\mu}}{\left(e^{\beta\omega} - 1\right)^2},
\end{equation}
whose high-temperature limit does not depend on the frequencies of the harmonic oscillator:
\begin{equation}
    \lim_{T\rightarrow\infty} C_v^{\text{(harm)}} = 3Nk_b.
\end{equation}
This limit is reached exponentially fast above the Debye temperature of the highest frequency (in \ch{CsSnI3} it is about \SI{160}{\per\centi\meter}, equivalent to \SI{230}{\kelvin}); therefore, in all the simulation reported, this term is constant and equal to the high temperature limit. 
The intrinsic anharmonic contribution instead is:
\begin{equation}
    C_v^{\text{(anharm)}} = T \sum_{\mu=1}^{3N} \frac{\partial S}{\partial\omega_\mu}\frac{\partial \omega_\mu}{\partial T} = \sum_{\mu=1}^{3N}\frac{\beta\omega_\mu e^{\beta\omega_\mu}}{\left(e^{\beta\omega} - 1\right)^2}\frac{\partial\omega_\mu}{\partial T} ;
\end{equation}
\begin{equation}
    C_v = C_v^{\text{(harm)}} + C_v^{\text{(anharm)}} = k_b\sum_{\mu=1}^{3N}\frac{\beta^2\omega_\mu^2e^{\beta\omega_\mu}}{\left(e^{\beta\omega} - 1\right)^2}\left(1 + \frac{T}{\omega_\mu}\frac{\partial \omega_\mu}{\partial T}\right).
\end{equation}
This expression gives the \eqname~\eqref{eq:heat:sscha} reported in the main text in the limit $T\rightarrow\infty$.

Notably, the frequencies that appear in this equation are the SSCHA auxiliary frequencies, which are positive definite by construction, and never go to 0 (not even at a phase transition).
Therefore, what causes the divergence of the specific heat at the phase transition is not the $1/\omega$ factor, but an inflection in the $\omega(T)$ curve. Thanks to this feature, the heat capacity remains well defined even below the transition, when the structure is metastable.

    \newpage
    \begin{figure}[p]
    \includegraphics[width=\textwidth]{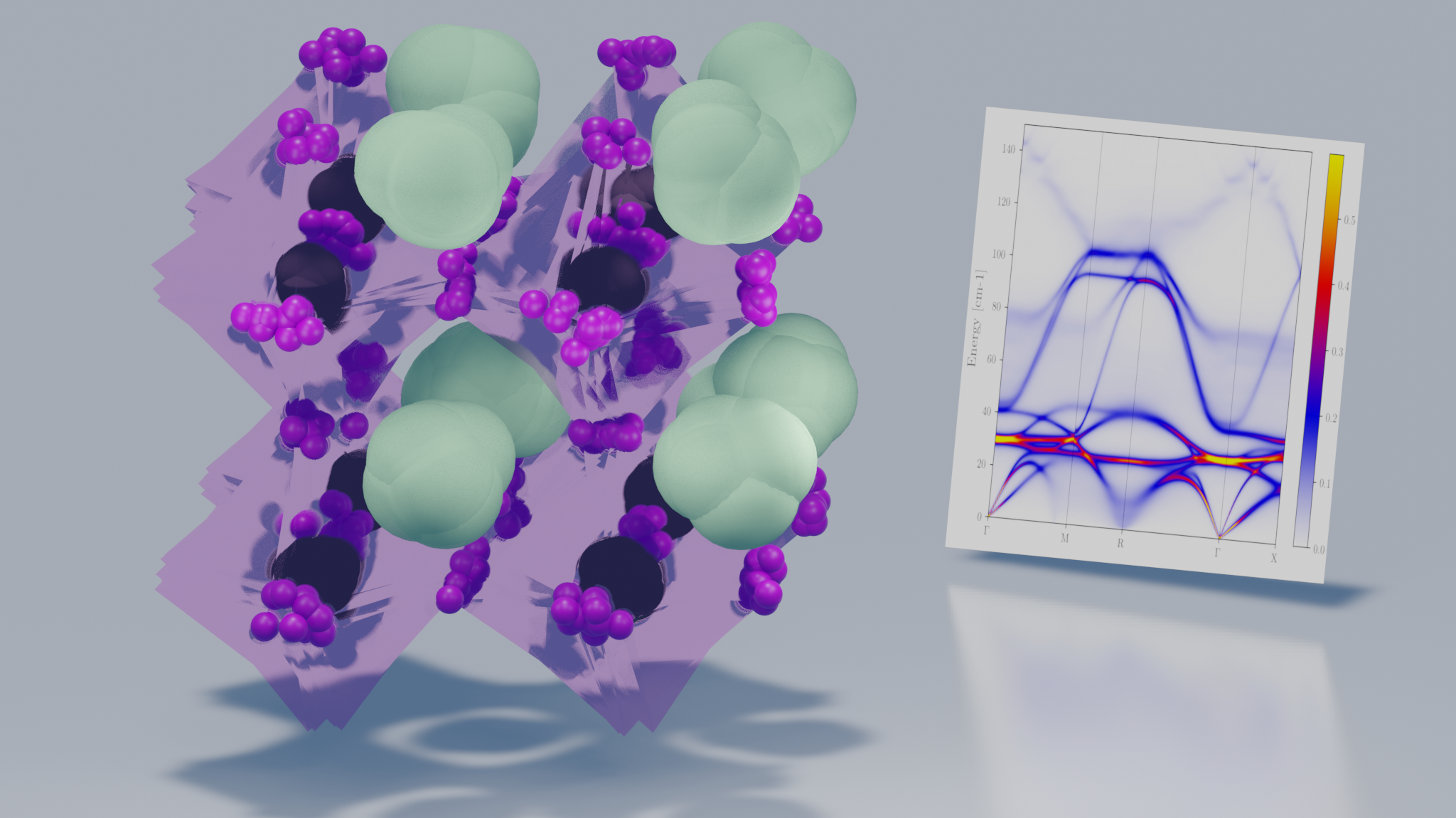}
    \caption{ToC figure; CsSnI3 perovskite structure in the cubic B-$\alpha$ phase, depicted with the quantum and thermal fluctuations. On the right side, the spectral functions of the anharmonic vibrations. Made with the software Blender 3.1.4}
    \end{figure}

\end{document}